\documentclass[traditabstract]{aa}
\usepackage{graphicx}
\usepackage{txfonts}
\usepackage{amssymb}
\usepackage{lscape}

\begin{document}

\def\mpc{h^{-1} {\rm{Mpc}}} 
\def\kpc{h^{-1} {\rm{kpc}}}
\newcommand{\mincir}{\raise
-2.truept\hbox{\rlap{\hbox{$\sim$}}\raise5.truept\hbox{$<$}\ }}
\newcommand{\magcir}{\raise
-2.truept\hbox{\rlap{\hbox{$\sim$}}\raise5.truept\hbox{$>$}\ }}

\title{The dichotomy of Seyfert 2 galaxies:\\ intrinsic differences and evolution}

\author{E. Koulouridis}

\institute{Institute for Astronomy \& Astrophysics, Space Applications \&
  Remote Sensing, 
National Observatory of Athens, Palaia Penteli 15236, Athens, Greece.}

\date{\today}

\abstract{
We present a study of the local environment ($\leq$200 $\kpc$) of 31 hidden
broad line region (HBLR) and 43 non-HBLR Seyfert 2 (Sy2) galaxies in the nearby
universe ($z\le$0.04). To compare our findings, we constructed two control samples
that match the redshift and the morphological type distribution of the HBLR and non-HBLR 
samples. We used the NED (NASA extragalactic database) to find all neighboring galaxies within a projected 
radius of 200 $\kpc$ around each galaxy, and a radial velocity difference $\delta u\le 500 km/s$. 
Using the digitized Schmidt survey plates (DSS) and/or the Sloan Digital Sky Survey (SDSS), when available, 
we confirmed that our sample of Seyfert companions is complete. 
We find that, within a projected radius of at least 150 $\kpc$ around each Seyfert, 
the fraction of non-HBLR Sy2 galaxies with a close companion is significantly higher than that of their
control sample, at the 96\% confidence level. Interestingly, the difference is due to the high frequency of
mergers in the non-HBLR sample, seven versus only one in the control sample, while they also present 
a high number of hosts with signs of peculiar morphology.  
In sharp contrast, the HBLR sample is consistent with its control sample. Furthermore,
the number of the HBLR host galaxies that present peculiar morphology, which probably implies
some level of interactions or merging in the past, is the lowest in all four galaxy samples.
Given that the HBLR Sy2 galaxies are essentially Seyfert 1 (Sy1) with their broad line region (BLR) hidden
because of the obscuring torus, while the non-HBLR Sy2 galaxies probably also include true Sy2s that lack the BLR as well as 
heavily obscured objects that prevent even the indirect detection of the BLR, 
our results are discussed within the context of an evolutionary sequence of activity 
triggered by close galaxy interactions and merging. 
We argue that the non-HBLR Sy2 galaxies may represent different stages of this sequence, possibly the beginning 
and the end of the nuclear activity.}


\titlerunning{On the dichotomy of Seyfert 2 galaxies}
\maketitle

\section{Introduction}

Polarized emission from the central engine of active galactic nuclei (AGN) can be produced by the
reflection of radiation from a scattering media.
Evaporated gas that originated in the inner surface of the obscuring matter near the
central black hole can escape the area and form such a mirror in a favorable
location where the light can be reflected toward the observer without being
absorbed. Thus, polarized light is a powerful tracer of AGN activity, otherwise
hidden due to obscuration along our line of sight (e.g., Krolik 1999).

Nearly thirty years ago, the first discovery by Miller \& Antonucci (1983) of
broad permitted emission lines and a clearly non-stellar continuum in the
polarized spectrum of the archetypal Seyfert 2 (Sy2), NGC 1068, was just the beginning
of numerous similar observations in a wide variety of galaxies. Ten years
later, the unification model (UM) of AGN  was formulated upon these observations
(Antonucci 1993). According to the UM, all Seyfert nuclei are intrinsically identical, while
the only cause of their different observational features is the 
orientation of an obscuring torus with respect to our line of
sight. A hidden broad line region (HBLR) was considered to be present in all Type II active galaxies,
visible to us only by the reflection of a fraction of the total emitted
radiation.

Nevertheless, polarization has also been a basis on which the unified scheme was
questioned, since Tran (2001, 2003) concluded that there is a non-HBLR Sy2
population significantly different from the corresponding HBLR Sy2 population.
He argued that, in contrast to the HBLR Sy2s, the true Sy2 AGN are
intrinsically less powerful and they cannot be fitted within the realm of the
UM. The same conclusion was reached in recent statistical studies (e.g., Tommasin
et al. 2010 and Wu et al. 2011). Many authors argue that the lack of a broad
line region (BLR) in the center of these Sy2 galaxies is luminosity and/or
accretion rate dependent (Nicastro 2000; Lumsden and Alexander 2001; Gu and
Huang 2002; Martocchia 2002; Panessa \& Bassani 2002; Tran 2003; Nicastro et al.
2003; Laor 2003; Czerny et al. 2004; Elitzur \& Shlosman 2006; Elitzur 2008; Elitzur and Ho 2009;
Marinucci 2012, Elitzur, Ho \& Trump 2014). Low luminosity can be the result of very low accretion rate and
the BLR may possibly be absent in such systems.

In addition, although widely accepted today, the UM cannot explain various
observed differences between Type I and Type II AGN. Many studies over the
years challenged its validity and proposed instead an evolutionary sequence
that links different types of activity (e.g., Hunt \& Malkan 1999;
Dultzin-Hacyan, 1999; Krongold et al. 2002; Levenson et al. 2001; Koulouridis et
al. 2006a, b, 2013). In particular, although the role of
interactions on induced activity is still an open issue (Koulouridis et al.
2006a,b, 2013 and references therein), most of the above studies seem to conclude that
the possible evolution of activity follows the path of interactions $\rightarrow$
enhanced star formation $\rightarrow$ Type II AGN $\rightarrow$ Type I AGN.
However, their Sy2 samples were never examined for the possible existence of
hidden broad lines since spectropolarimetric observations are time-consuming
and are feasible only with large telescopes. 

Koulouridis et al. (2006a) compared the environment of two samples of Seyfert 1 (Sy1) and
Sy2 galaxies to that of two well-defined control samples,
concluding that the Sy2 sample shows an excess of close companions, while the
Sy1 sample did not. In addition, Koulouridis et al. (2013) showed that
the neighbors of Sy2 galaxies are systematically more ionized than the
neighbors of Sy1 galaxies, a fact that indicates differences in metallicity, stellar
mass, and star-formation history between the samples. In the current study we focus solely on Sy2
galaxies by investigating the environment of the biggest compiled HBLR and
non-HBLR samples to date, in order to discover possible differences between the
two that can provide us with additional clues about the nature of these objects. 

We describe our samples and methodology in Sect. 2, our results in Sect. 3, and our conclusions and discussion in Sect. 4. 
Throughout this paper we use $H_0=73$ km/s/Mpc, $\Omega_m=0.27$, and
$\Omega_{\Lambda}=0.73$.

\section{Sample selection \& methodology}

\subsection{On the non-HBLR population}
Although the existence of non-HBLR Sy2s can be succesfully explained by the true Sy2 interpretation, 
this is not the only solution to the lack of observed broad emission lines (see also Antonucci 2012). 
As indicated by previous studies (e.g., Wu et al. 2011),
in spite of the overal differences, the properties of a fraction of the non-HBLR Sy2s are very similar to the properties of the HBLR
Sy2s. Therefore, we argue that we can divide the non-HBLR Sy2s into three main categories based on their obscuration:
\begin{enumerate}
\item true Sy2s, that do not present broad emission lines in their spectra, but 
at the same time are not obscured (e.g., Panessa \& Bassani 2002; Akylas \& Georgantopoulos 2009); 
\item heavily obscured, so that the presence of a possible BLR cannot be detected in any way;
\item mildly obsured, so that the BLR was possibly not observed because of other limitations, 
e.g., observational flux limit, host galaxy obscuration, bad orientation, or total
lack of the scattering matterial that produces the polarized broad lines (if this is not the same as the obscuring matterial). 
\end{enumerate}
Only in the first case can we be almost certain that the object does indeed lack the BLR and it is intrisically different
from a broad line Seyfert. In the other two cases the BLR can either be present or not.
Therefore, we should clearly state once more that the current study investigates two samples of HBLR and non-HBLR Sy2s, without
making any other assumption on the nature of these objects. The discovery of any differences in the environment 
of these two samples may indicate intrinsic differences between the two populations that do not necessarily apply to all 
objects individualy.

\subsection{Sample selection}
The original HBLR and non-HBLR Sy2 galaxy samples can be found in Wu et al.
(2011). From this sample we excluded all galaxies with $z\ge 0.04$ since 
above this limit their morphological type is usually undefined and 
the number of probable neighbors with no redshift becomes 
very large for our statistics. In addition, we find that even their classification 
as Sy2 is uncertain, let alone the detection of polarized broad emission lines.
The detection is dubious because in order to calculate the polarized
fraction of the reflected light, the stellar contribution must be subtracted,
which is more difficult for smaller objects where more starlight is included in
the observed spectrum (Krolik 1999). From the original catalogue we also excluded 
the 18 Sy2s that have no spectropolarimetric observations but were classified using 
other criteria by Wang \& Zhang (2007). Finally, a small number of faint 
galaxies were excluded independently of their redshift.
\begin{figure}
\resizebox{14cm}{14cm}{\includegraphics{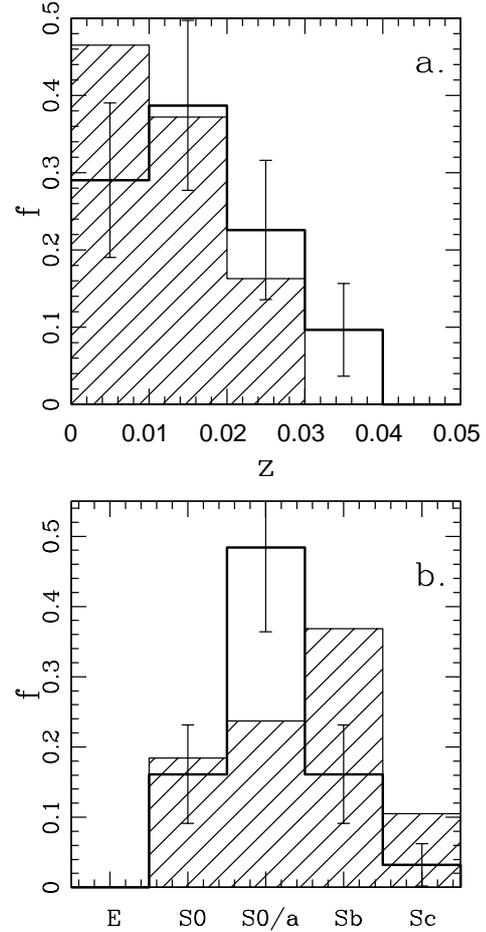}}
\caption{Redshift (panel a.) and Morphological type (panel b.)
distribution of the Sy2 samples. The solid line defines the
HBLR Sy2 population, while the hatched histogram the non-HBLR. Uncertainties are
1$\sigma$ Poissonian errors and they are plotted only for the HBLR Sy2s for clarity.}
\end{figure}
\begin{figure*}
\resizebox{18cm}{9cm}{\includegraphics{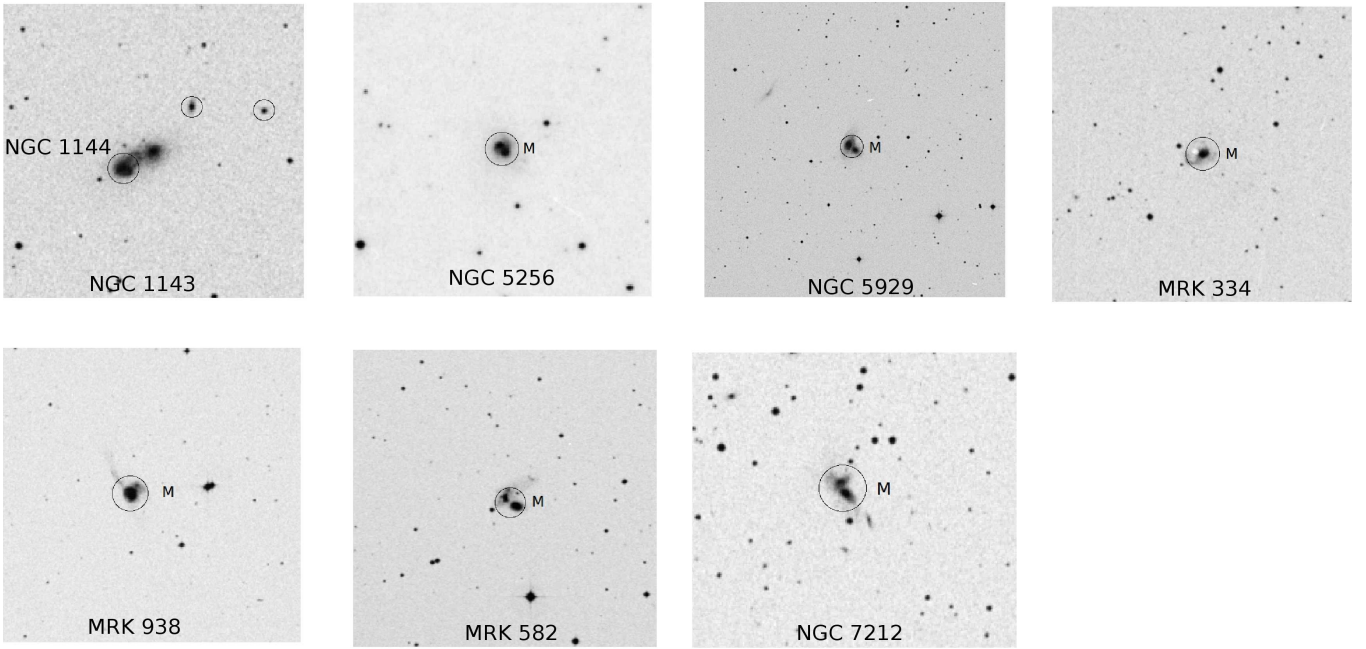}}
\caption{Images of the merging galaxies in our samples.}
\end{figure*}

The redshift and the morphological type distribution of the two Seyfert samples are 
presented in Fig. 1. The distributions do differ, especially the morphological, although their differences are not
statistically significant. An  
interesting trend is that the HBLR Sy2s are hosted by earlier-type galaxies than the non-HBLR.
This trend is already known for the Sy1 and Sy2 hosts, where the former show the 
same behavior as the HBLR Sy2 hosts. This supports the interpretation that 
the HBLR Sy2s are indeed Sy1s with their broad line region obscured by a dusty torus,
but reflected by a scattering surface located over the torus. On the other hand, this trend 
already implies intrinsic differences between the two Sy2 types. Because of these
differences in the distributions, we do not proceed with a direct comparison of the environment of our two samples.  
The slightly different redshift distribution may introduce a bias against the detection of HBLR Sy2 neighbors because  
the higher the redshift, the higher the probability of fainter neighbors with unknown redshift. On the 
other hand, the different morphological type distributions may lead to the opposite direction, since it is well
known that early-type galaxies are more clustered than late-types. Thus, we choose to build 
two control samples that have the same redshift and morphological type distribution with the two Sy2 samples, 
respectively. Any comparison will be made between each Sy2 and its control sample and the conclusions 
will be drawn from there.

The control samples were randomly built from the NGC, MRK, and IC databases.
All objects that had any reference of being active (AGN or LINER) were excluded and 
replaced, and finally the samples were refined so as to match the redshift and morphological 
type distribution of the HBLR and non-HBLR samples. To have a homogeneous 
morphological classification we chose to use the types listed in the Third Reference Catalogue
of bright galaxies (RC3). We should note that peculiarity was not taken into consideration when 
constructing the control samples as this could bias our results, 
i.e., a peculiar Sa galaxy is treated like an Sa for our purposes.
In addition, galaxies classified solely as peculiars were not included in the morphological type 
distribution matching at all. In more detail, in the control samples we did 
not attempt to include the same number of peculiar galaxies that we found in the Seyfert samples. 
The reason is that peculiar galaxies may have undergone a recent merger or may be still 
strongly interacting with a close companion. Forcing the same number of peculiar galaxies in the control 
sample could artificially enhance the number of interacting galaxies.
\begin{figure*}
\resizebox{18cm}{9cm}{\includegraphics{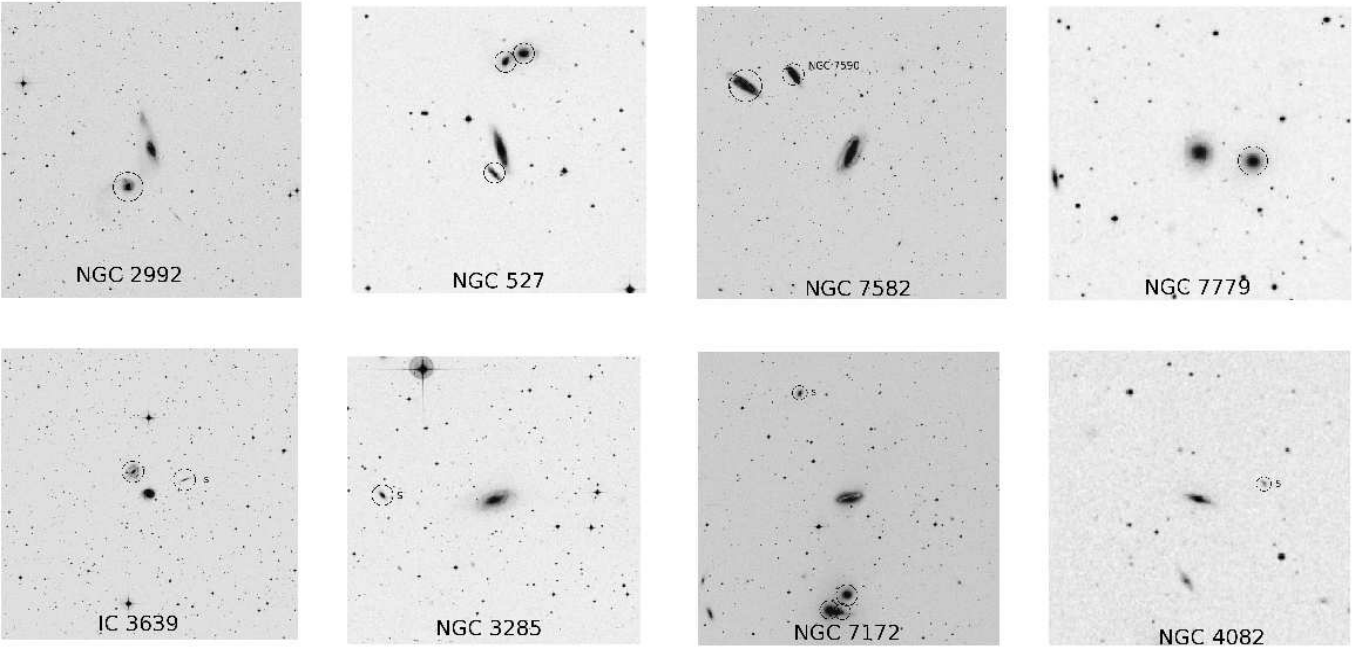}}
\caption{Examples of Seyferts and control sample galaxies with companions. Small companions 
are marked with the letter s. In the case of NGC 7582, another non-HBLR Sy2, NGC 7590 is one 
of the close neighbors.}
\end{figure*}

\subsection{Methodology}
In order to identify possible neighbors around each Sy2 and control sample galaxy, 
we made use of the automated search of the NASA/IPAC Extragalactic 
Database. In the current study we do not discriminate between galaxy pair and satellite 
galaxies. All galaxies that meet the qualifying criteria described below are considered as neighbors of 
the investigated galaxy:
\begin{itemize}
\item galaxies within a projected 
radius of 200 $\kpc$ from the AGN or the control galaxy, 
\item galaxies with a radial velocity difference of $\delta u\le 500 \;km/s$. 
\end{itemize}
Even though there is no general consensus on the maximum radial separation
of a galaxy pair, most of the recent studies use a search radius
between 20$\kpc$ (e.g., Patton et al. 2005) and 200$\kpc$ (e.g.,
Focardi et al. 2006; see also relevant discussion in Deng et al. 2008). 
We chose the limit of 200$\kpc$ considering that it is a
reasonable distance for a satellite galaxy in a massive halo
(e.g., Bahcall et al. 1995; Zaritsky et al. 1997). Distances were estimated 
taking into account the local velocity field, which includes the effects of 
Virgo, Great Attractor and Shapley, for the standard $\rm\Lambda$CDM cosmology 
($\Omega_m$=0.27, $\Omega_\Lambda$=0.73)
Since, however, NED is not complete in any way, we also visually inspected 
all DSS or SDSS images (when available) to discover any other neighbor candidates with no 
listed redshift. Only four possible neighbours with no available spectroscopic redshift were found, two of the HBLR and two of their
control sample. However, based on SDSS photometric redshifts that were available for all four objects, we chose to exclude them from 
the analysis since they were incompatible with the sample galaxy redshift, even within the errors. We note that the inclusion 
of these four objects would not alter the results.
Finally, we classified the companion galaxies
based on their magnitude difference ($\Delta m$) with the sample galaxy
by using a common blue magnitude, mostly from the RC3 or the SDSS 
database, i.e., the neighbors with $2<\Delta m\le 3$ were classified as small 
and were treated separately, while those with $\Delta m> 3$ were 
not included in this analysis. In addition, above 100$\kpc$ small neighbors were not included. 
Mergers were regarded as galaxies in strong interaction regardless of the magnitude difference of the 
pre-merging galaxies, which is unknown in many cases. A wide range of cases were considered as mergers, 
from the case of two clearly separated but nevertheless very close galaxies with clear signs of strong interactions 
(e.g., NGC 1143 and NGC 1144) to the case of post-mergers with peculiar morphology, only if 
reported as such in the literature (e.g., MRK 334). The images of all the merger galaxies 
can be found in Fig. 2. Finally, we note that MRK 1039 and MRK 1066 are reported as galaxies with 
double nucleus. We chose to include them as mergers (probably in an advanced stage).

Our final Seyfert samples consist of 43 non-HBLR, 31 HBLR Sy2s, and their control samples 
(a total of 148 galaxies). The galaxy redshift, the morphological type and 
the projected distance to their neighbors are given in
Tables 1 and 2. Our search radius of 200 $\kpc$ was divided into four bins of 
50 $\kpc$ and each neighbor was placed accordingly. In the current study we are mostly
interested in addressing the following two questions: 
\newline
a. Is there any significant difference between the fraction of Seyferts and the fraction of 
their respective control sample galaxies that have at least one neighbor within a given radius?
\newline
b. Is there any significant difference between the density of the environment where the Seyferts 
and their respective control sample are found?
\newline
A positive answer to either question will imply that there are intrinsic differences between the
investigated samples, and that the environment probably plays a leading role in the creation of these differences.
The answer to the first question will show us if interaction with a close neighbor can play the role of the triggering mechanism  
that leads to the detected differences, while to the second will provide us with information about the general 
local environment. We should stress that in the present study we are not interested 
in the large-scale environment of our sample galaxies. This will probably not differ between each Seyfert sample 
and its respective control sample, since they were selected to have the same redshift and, especially, 
morphological type distribution. However, the local environment may vary, as shown in Koulouridis et al. (2006),
independently of the density of the large-scale environment. Even the presence of a single neighbor 
of comparable mass, within a certain radius, may introduce the necessary conditions that would lead to the
detected differences.

%\linespread{1}

\section{Results}

In Tables 1 and 2 we listed all our galaxies, active and control, and their
close companions divided into four bins depending on their radial separation.
With the symbol ``x'' we mark all neighbors that are no more than two magnitudes fainter than the 
galaxy in question, while with ``s'' the ones that are between two and three magnitudes 
fainter. The selection of these specific limits is not random, but was decided after 
visual inspection of the neighbors. Some characteristic examples can be found in Fig. 3,
where we can see that in contrast to neighbors with $\delta m<2$, the ones above 
this limit are becoming rather small in comparison and their capability of producing 
any sufficient interactions can be questioned. In addition, in the case of higher redshift galaxies 
and those in regions with significant star contamination, faint galaxies may be missed, although 
these effects will probably affect equally the Seyfert samples and their control. For these 
reasons, although listed in the tables, small galaxies were not considered when we extracted our main 
results, and their possible role is only discussed briefly. Given their ambiguous role, even in
small radial distances, above 100 $\kpc$ small neighbors are not listed at all. Except for these neighbors,
visual investigation also revealed a small number of companions, of comparable size to the central ones, but with 
no redshift information. These objects are listed within parentheses and are included in our analysis, although
they would not play any significant role if they were excluded. Finally, for a small number of cases, the decision whether
the companion is large or small had to be made after visual investigation because of lack of data. These cases are 
marked with a star.

To draw our conclusions we will refer to the results of a number of statistical tests.
First, we ran the Fisher's exact test for a 2$\times$2 contingency table to compare 
the fraction of Seyferts and the fraction of their respective control sample galaxies 
that have at least one neighbor within a given radius (question a. in the methodology). 
Each row of the contingency table is one of the two samples (Seyfert and respective control sample), 
while the input in the two columns is the number of galaxies
that have a companion (Col. 1) and do not have one (Col. 2) within a certain radius. Therefore, the sum of the two columns of each
row gives us the total number of galaxies in each sample. The results of the tests are listed in the 
last row of Tables 1 and 2. We list the results of four tests in each table, depending 
on the radial distance we chose to search for neighbors. We always consider previous radial 
bins in the total, i.e., the second value in Table 1 ($P_{null}$=0.7723) refers to the comparison 
of the total number of HBLR Sy2 and control sample galaxies that have companions 
within 100 $\kpc$ (not between 50 and 100 $\kpc$); $P_{null}$ is the probability that the two samples 
are drawn from the same parent population. For the HBLR Sy2 and its control sample, independent 
of the search radius, the results suggest that the null hypothesis cannot be rejected at any significant 
statistical level, and thus the two samples are probably drawn from the same parent population. In sharp 
contrast, the non-HBLR and its control sample do indeed differ at a high statistical level in the first three bins ($>$96\%),
and at a moderate level in the fourth. We note that for the non-HBLR analysis, the one-sided test was used, since
we already suspected from our previous works (Koulouridis et al. 2006a, b, 2013)
that a real Seyfert 2 population will present more companions than their control sample galaxies. 
We note, however, that the two-sided test results are two times the one-sided results, 
and therefore by using it we would reach qualitatively the same conclusions, 
although after the inclusion of the last bin there would be no statistically significant difference at any level
between the two samples.

Independently of the results of the previous paragraph, to compare the density of the environment that the Seyferts 
and their respective control sample are embedded in (question b. in the methodology), 
we calculated the mean 
number of galaxies that are found within a 200 $\kpc$ radius around all the galaxies
of each sample. Then, we tested how 
statistically significant the difference of the means is, by calculating the error $\sigma$ of these 
differences with the formula
\[\sigma=\sqrt{\frac{(n_1-1)S_1^2+(n_2-1)S_2^2}{n_1+n_2-2}\left[\frac{1}{n_1}+\frac{1}{n_2}\right]},\]
where $n_1$ and $n_2$ are the number of sample galaxies and $S_1$ and $S_2$ the standard 
deviation. In the case of the HBLR Sy2s and their control sample $\sigma=0.228$, and the difference 
of the means is $\delta \overline{x}=0.548$, where $\overline{x}$ is the mean number of companions in each sample. 
These results show 
that the environments of the two populations are different at the 97.8\% level, with the Sy2s showing a 
preference for less dense environments. Similarly, for the 
non-HBLR and its control sample the respective values are $\delta \overline{x}=0.512$ with $\sigma=0.218$
and again the difference of the means is statistically highly significant, at the 98.9\% level. However,
in this case, the preference of the Sy2 population is for the denser environments.

Finally, we attempted to compare more directly the two Sy2 populations by calculating the overdensity
of companions around each host. This overdensity measure is given by the formula
\[\delta=\frac{x-\overline{x_c}}{\overline{x_c}},\]
where x is the number of neighbors around each Sy2 host, and $\overline{x_c}$ is the mean 
number of companions derived from the control sample. In this way we took into consideration the 
information of the control sample, while at the same time we only had two sets of numbers to compare,
the overdensities of the HBLR and the non-HBLR Sy2s. We ran a Kolmogorov-Smirnov test and we concluded 
that the overdensities differ at a very high statistical level ($>$99.9\%).

We note that small neighbors were not considered in the above statistical analyses for the reasons 
we described in the methodology. However, we mention that this could affect mostly the Fisher's exact test
for the non-HBLR and its control sample because there is a number of small galaxies to be found within 
the first 50 $\kpc$ around the control galaxies. None of the remaining results would change significantly. We 
also note that the control galaxy NGC 4488 is located within a galaxy cluster and, as noted in Table 2, 
a large number of companions can be found in its third and fourth bin. However, for the statistical analyses we chose to consider 
only two companions in each bin, because otherwise we believe that the results would be biased.

Interestingly, we note the relatively large number of mergers and peculiar host morphologies in the non-HBLR sample
in comparison to their control sample. In particular, the merger fraction is significantly higher when compared to any of our samples.
These facts corroborate with our findings that the non-HBLR host galaxies reside 
in denser environments when compared with their control sample and a significantly larger fraction of them
presents at least one companion within 200 $\kpc$. In many cases, by visual investigation of the DSS and/or the SDSS images,
 we can clearly identify the companion galaxy or merging as the reason for the peculiar morphology. 
However, this is only visible in the case of close pairs ($D<30$ $\kpc$), while for galaxies with no companion, the 
peculiar morphology probably suggests past interactions/merging or even false classification.
On the other hand, the HBLR Sy2 sample has only three hosts with signs of peculiar morphology, less than half of its control
sample, which again agrees with their preference for less dense environments and fewer interactions.

\section{Discussion and conclusions}
Investigating the environments of two samples of HBLR and non-HBLR Sy2s,
we reached the conclusion that there is a
statistically significant difference between the two types of
Sy2 galaxies and their respective control samples. The non-HBLR population have 
neighbors more frequently than their control sample galaxies, within the specified spatial and velocity limits.
Also, they are found more often in denser close environments than galaxies of the same 
morphological type, and the frequency of merging and peculiar morphologies is also relatively high.
On the contrary, HBLR Sy2s are found in less dense environments than their control sample, and the
frequency of merging or any sign of interactions is also low. In addition, 
their fraction with at least one neighbor agrees with the control galaxies, in all search radii.
These results are also in agreement with our previous studies on Seyfert galaxies and they indicate
the similarities between HBLR Sy2 and Sy1 galaxies, supporting the view that in all probability they are intrinsically 
the same objects. However, non-HBLR Sy2 galaxies seem to differ significantly, 
although their nature is still a matter of debate. Therefore, in the light of the current results, we will discuss 
probable mechanisms responsible for the observed and possibly intrinsic differences.

There is strong evidence that the dusty obscuring torus in low luminosity AGN is
absent or is thinner than expected in higher luminosities (e.g., Whysong \&
Antonucci 2004; Elitzur \& Shlosman 2006; Perlman et al. 2007; van der Wolk et
al. 2010). Accordingly, all low luminosity AGN should have been Type I sources,
which of course is not the case. The only reasonable explanation to this problem
is the additional absence of the BLR in such systems. As we have noted earlier,
some authors (e.g., Nicastro 2000; Nicastro et al. 2003; Bian et al. 2007; Marinucci et al. 2012; Elitzur, Ho \& Trump 2014)
presented arguments that below a specific accretion rate of material into the
black hole, and therefore at lower luminosities, the BLR might be absent.
Elitzur \& Ho (2009), using data from nearby bright AGN, concluded that the BLR
disappears at bolometric luminosities lower than $5 \times 10^{39} (M/10^7 M_{\sun})^{2/3}
\rm erg\; s^{-1}$, where $M$ is the mass of the black hole. They also argued
that the quenching of the BLR, and the disappearance of the torus can occur
either simultaneously or in sequence, with decreasing black hole accretion rate and
luminosity. Thus, a possible scenario would be that non-HBLR Sy2 AGN are objects 
lacking the BLR and possibly the torus. 
Bian et al. (2007) and Wu et al. (2011) separates their  non-HBLR sample into luminous and less luminous 
using the $\log L_{\rm [OIII]}<41$ limit, while Marinucci et al. (2012), argued that true Sy2s can be found below the 
bolometric luminosity limit $\log L_{bol}=43.9$. We should note that in Marinucci et al. (2012) 
they derived the bolometric luminosity from the X-ray and the 
[OIV] luminosity and concluded that $L_{[OIII]}$ is not as reliable (see also relevant discussion in Elitzur 2012).

An alternative scenario is that heavy obscuration in non-HBLR Sy2 does not 
allow the detection of the BLR even in the polarized spectrum.
Marinucci et al. (2012) concluded that 64\% of
their compton-thick non-HBLR Sy2s exhibit higher accretion rates than the
threshold clearly separating the two Sy2 classes. They attributed this
discrepancy to heavy absorption along our line of sight, preventing the
detection of the actual BLR in their nuclei. Evidently, merging systems constitute
a class of extragalactic objects where heavy obscuration occurs (e.g., Hopkins et
al. 2008). The merging process may also lead to rapid black hole growth, giving
birth to a heavily absorbed and possibly compton-thick AGN. Thus, we could
presume that a fraction of our non-HBLR mergers, if not all of them, might actually be
BLR AGN galaxies, where the large concentration of gas and dust prohibits even
the indirect detection of the broad line emission (e.g., Shu et al. 2007). However, other studies 
concluded that there is no evidence that non-HBLR Sy2s are more obscured than their HBLR peers (Tran 2003; Yu 2005; Wu 2011), while 
totally unobscured low-luminosity non-HBLR Sy2s were detected via investigation 
of their X-ray properties (e.g., Panessa \& Bassani 2002; Akylas \& Georgantopoulos 2009). The total population of non-HBLR Sy2s
is probably a mixture of objects with low accretion rate and/or high obscuration. 

Both scenarios agree with the interpretation of our current results by the
evolutionary scheme proposed by Krongold et al. (2002) and supported later by
Koulouridis et al. (2006a, b, 2013), according to which, interaction with a comparable
sized galaxy can drive molecular clouds toward the nucleus and trigger an
evolutionary sequence, going from enhanced star formation to obscured Type II and
finally to Type I activity. If the first scenario is valid, it is to be expected that during the initial stage
of the interaction the accretion rate of the central black hole would be low
and there would be neither a BLR nor a torus. 
In addition, as already discussed, heavy absorption caused by the interaction
may also prevent the detection of the possibly existing BLR during the first stage of the 
AGN cycle. Consequently, the first stage of nuclear activity should be a non-HBLR narrow line AGN.
Evidently, 71\% of the non-HBLR Sy2 galaxies, with at least one neighbor within 100$\kpc$ and/or
peculiar morphology, have high obscuration ($N_H>10^{24}$) and/or low luminosity ($\log L_{\rm [OIII]}<41$; 
criterion by Wu et al. 2011). For another 19\% we do not have the information, 
while only 10\% of them are reported as luminous and at the same time unabsorbed sources. Although these fractions
may vary depending on the BLR-disappearance luminosity limit, we should mention than more than half
of our objects have column densities that characterize Compton-thick AGNs. In addition, 
our non-HBLR sample contains only ten galaxies (23\%) with $\log L_{\rm [OIII]}<41$, which can be considered as low luminosity, 
while all HBLR Sy2s are above that limit. 
However, this classification should not be considered explicit.
We argue that the lack of the BLR is not due to low luminosity per se, but rather to their 
being at the start of the activity duty cycle, which renders them less powerful 
than the ones that already have formed a BLR. Therefore, the actual value of the luminosity 
is not so relevant for the comparison, but instead the fact that as a whole the non-HBLR population
would be less luminous than the respective HBLR. Indeed, a Kolmogorov-Smirnov test indicates
that the luminosity distributions of our two Seyfert samples are significantly different at the 98.7\% level, with the HBLR Sy2s being 
shifted toward higher luminosities. In addition, Ho et al. (2012)
and Miniutti et al. (2013) report AGNs with very high Eddington ratios, but very low black
hole masses and no broad lines (see also Wang et al. 2012). 
Because of their high Eddington ratios, both objects greatly exceed
the luminosity limits above which, previous studies argue, the 
BLR should be present, and provide observational 
evidence that true Sy2s can have higher luminosities.

The accretion rate and luminosity increase can generate the BLR and also anticipate 
the heavy obscuration, leading to the
HBLR-Sy2 and finally Sy1 phase. However, the time needed for Type I activity to
appear should be larger than the timescale for an unbound companion to escape
from the close environment, or comparable to the timescale needed for an evolved
merger ($\sim 1$Gyr, see Krongold et al. 2002). This delay is a possible
explanation for the lack of close neighbors around the HBLR Sy2 and Sy1
galaxies (see Koulouridis et al 2006a, 2013) and for the earlier-type morphologies of
their hosts. If the evolutionary scenario is valid, unobscured Type I AGN can only
exist after the dissipation of the obscuring media and the strangulation of the
star forming activity by the AGN feedback (Krongold et al. 2007, 2009, Hopkins \&
Elvis 2010; see also the disk wind scenario in Elitzur \& Shlosman 2006). 
\begin{figure}
\resizebox{8cm}{8cm}{\includegraphics{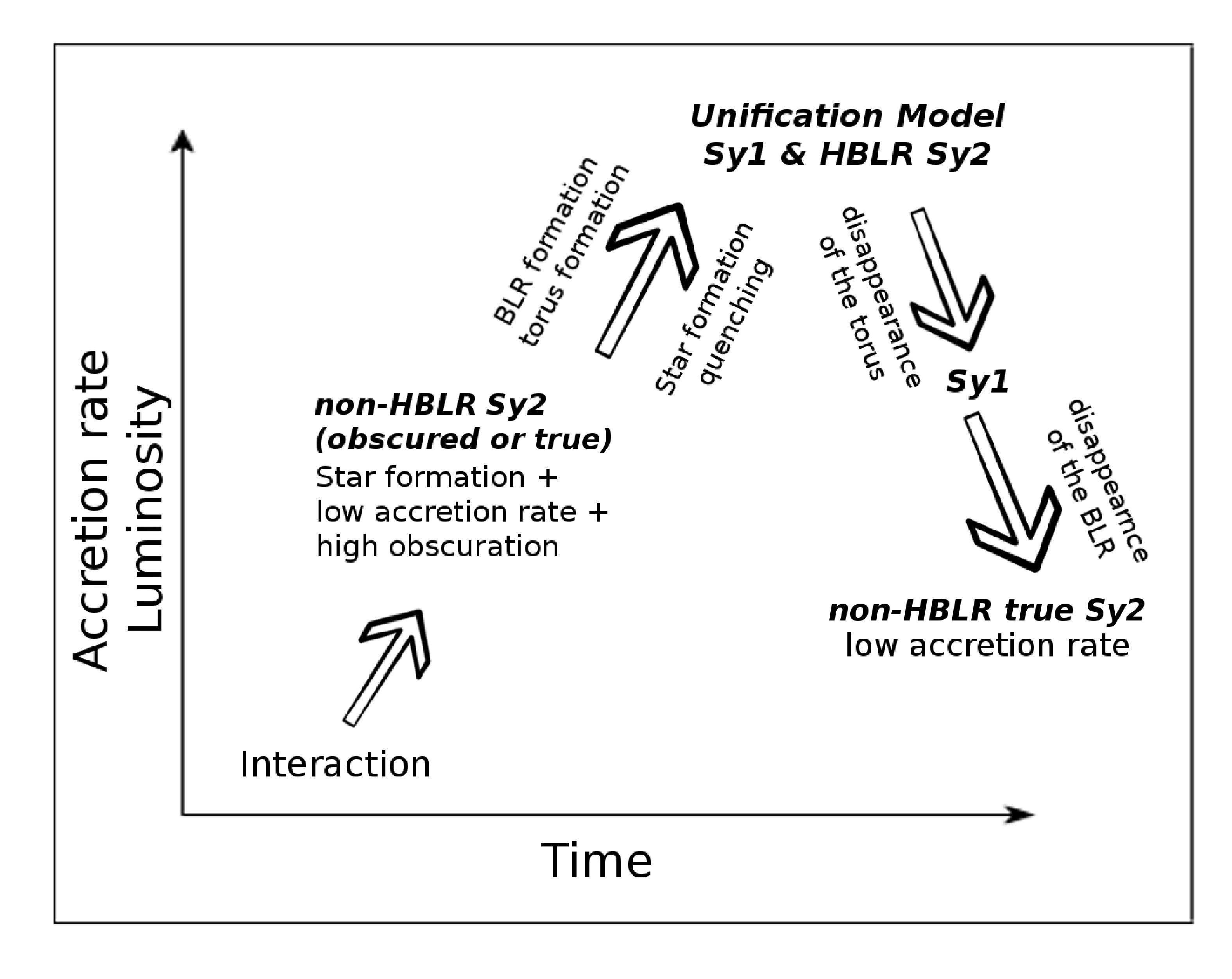}}
\caption{Qualitative description of the Seyfert evolutionary scheme.}
\end{figure}

On the other hand, about half of the non-HBLR Sy2s seem isolated and undisturbed.
These objects do not seem to be either just triggered or heavily 
obscured by recent close interactions.
However, if the low accretion rate scenario is valid, 
one would expect that AGN should also lose their BLR at the end of the AGN duty cycle,
as the accretion rate drops below a critical value (e.g., Bian et al. 2007; Elitzur \& Ho 2009; Elitzur, Ho \& Trump 2014).
Denney et al. (2014) argue that NGC 590 is an example of a Seyfert 1 that changed
to Type 1.9 in less than 40 years, and that this is due to a significant luminosity, and therefore
accretion rate, decrease. However, the morphological type distribution of the non-HBLR Sy2 host galaxies
(as we saw earlier in Fig. 1) peaks at even later-type spirals than the
corresponding distribution of the HBLR sample, although the difference is not 
statistically significant. This implies that the majority of these galaxies are
still unevolved. Even though the morphology of such a galaxy could
oscillate from late to early types up to four times, depending strongly on 
the environment and cold gas availability (Bournaud \& Combes 2002), the older
stellar population of the HBLR Sy2 and Sy1 hosts, compared to the non-HBLR,
reported by Wu et al. (2011), provides some evidence that non-HBLR Sy2 galaxies
probably precede the broad line phase (see however, Yu et al. 2013). However, although we cannot conclude positively
about this scenario, it agrees perfectly well with the final stage of the evolutionary 
scenario.

Considering all the above, HBLR and non-HBLR Sy2s can be fitted within the AGN evolutionary scheme as described 
qualitatively in Fig. 4 (see also Fig. 17 of Wang \& Zhang (2007) for 
a similar evolutionary scheme). The appearance of an unobscured Sy1 before the 
true Sy2 is also predicted by the disk wind scenario by Elitzur \&
Shlosman (2006). From the conclusion 
that the creation/disappearance of the BLR follows the increase/decrease of the accretion rate, emanates a logical prediction: at the stage of the
increase, the BLR is detectable when the accretion rate has already reached a relatively high level, while at the stage of decrease the BLR
is becoming non detectable when the accretion rate has already reached a relatively low level. Thus, the accretion rate-luminosity limit
for the detection of the BLR should be higher at the stage of increasing accretion rate and lower at the decreasing stage. This is
the reason why we plot in Fig. 4 the true Sy2s with different luminosity and accretion rate levels.

Alternatively, it also seems possible that the true Sy2 AGN is a stand-alone
phenomenon, caused by minor merging events or by secular evolution, probably
initiated or re-inforced by galaxy interactions (e.g., Combes 2011). Ho (2009) showed that
the low accretion rates can be supplied through local mass loss from evolved stars and Bondi 
accretion of hot gas, without any need for additional fueling
mechanisms. No activity evolution is expected if these
mechanisms cannot provide the required accretion rate to power the formation
of the BLR.

In a nutshell, the current and previous studies showed that at least a fraction of non-HBLR Sy2s
are probably intrinsically different from HBLR Sy2s, which in turn are probably 
obscured Sy1s. We argue that the non-detection of their BLR can be explained by
the intrinsic lack of it, because of the low accretion rate of gas and dust 
onto the super-massive black hole, or alternatively, by heavy obscuration that can successfully cloak
the BLR. We also argue that the
existence of true Sy2 can be fitted nicely within an evolutionary scheme, 
where a low accretion rate is predicted at the beginning and the end of the Seyfert 
duty cycle. Nevertheless, we cannot rule out the possibility that some HBLR Sy2s
could also be created by minor disturbances or even secular processes and that they turn off 
without any further evolution to other Seyfert types.

\acknowledgements
I would like to thank M. Plionis and I. Georgantopoulos for discussions and
useful suggestions. The research project (PE-1145) is implemented within the framework 
of the Action «Supporting Postdoctoral Researchers» of the Operational Program 
"Education and Lifelong Learning" (Action’s Beneficiary: General Secretariat for Research and Technology), 
and is co-financed by the European Social Fund (ESF) and the Greek State.
This research has made use of the NASA/IPAC Extragalactic Database (NED) which
is operated by the Jet Propulsion Laboratory, California Institute of
Technology, under contract with the National Aeronautics and Space
Administration. Funding for SDSS-III has been provided by the Alfred P. Sloan Foundation, the
Participating Institutions, the National Science Foundation, and the U.S.
Department of Energy Office of Science. The SDSS-III web site is
http://www.sdss3.org/.

\begin{table*}
\begin{minipage}{175mm}
\caption{HBLR Sy2 and control sample}
\tabcolsep 5pt
\begin{tabular}{lccccrc|lccccrcc}
{\em Name} &{\em 50} & {\em 100}&{\em 150}&{\em 200}&  {\em T$\;\;\;$}&{\em z} &
{\em Name} &{\em 50} & {\em 100}&{\em 150}&{\em 200}&  {\em T$\;\;\;$}&{\em z} &\\
{\em (1)} &{\em (2)} & {\em (3)}&{\em (4)}&{\em (5)}&  {\em (6)$\;\;$}&  {\em (7)}&
{\em (1)} &{\em (2)} & {\em (3)}&{\em (4)}&{\em (5)}&  {\em (6)$\;\;$}&  {\em (7)}\\
\hline
&&&&&&&&&&&&\\
F00317--2142 &s&&&&4&0.0268&MRK 133&x&&x&&4--pec&0.0068\\
F02581--1136 &&&&&1--pec&0.0299&MRK 179&&&x&&5&0.0111\\
IC 3639      &xs&&&&4&0.0109&MRK 449&&&x&x&1&0.0038\\ 
IC 5063      &&&&x&-1&0.0113&MRK 575&&&&&1&0.0183\\
MCG-03-34-64 &&&&&0&0.0165&MRK 582&M&&&&pec&0.0186\\
MCG-03-58-07 &&&&&0&0.0315&MRK 1363&&&&&1-pec&0.0091\\
MCG-05-23-16 &&&&&-1&0.0085&NGC 527&x&xx&&&0&0.0193\\
MRK 3        &&&x&&-1&0.0135&NGC 634&&&&&1&0.0164\\
MRK 78       &&&&&1&0.0370&NGC 691&&&x&x&4&0.0089\\
MRK 348      &&&&&1&0.0150&NGC 1168&&&&x&3&0.0254\\
MRK 477      &&x&&&-1&0.0377&NGC 2221&x&&&&1-pec&0.0084\\
MRK 1210     &&&&&1&0.0135&NGC 3092&x&xx&x&&-1&0.0197\\
NGC 424      &&&&&1&0.0118&NGC 3182&&&&&1&0.0071\\
NGC 513      &s$\star$&&&&4&0.0195&NGC 3285&&s&&x&1-pec&0.0113\\
NGC 591      &&&&&1&0.0151&NGC 4488&&&x*&x*&0-pec&0.0032\\
NGC 788      &&&&&1&0.0136&NGC 4608&&x&&&-1&0.0062\\
NGC 1068     &&&&x&3&0.0038&NGC 5352&&&&&-1&0.0266\\
NGC 2110     &&s$\star$&&&-1&0.0078&NGC 5607&&&&&pec&0.0253\\
NGC 2273     &&&&&1&0.0061&NGC 3179&&&&&-1&0.0242\\
NGC 2992     &x&&&&1--pec&0.0077&NGC 6660&&&&x&0&0.0141\\
NGC 3081     &&&&&0&0.0080&NGC 7312&&&&&3&0.0277\\
NGC 4388     &&&&&3&0.0084&NGC 7415&&&&&2&0.0399\\
NGC 4507     &&&&&3&0.0118&NGC 2375&&&&&3&0.0262\\
NGC 5252     &&&&&-1&0.0230&NGC 1486&&&&&4-pec&0.0248\\
NGC 5506     &x&&&&1--pec&0.0062&NGC 7272&&&&&1&0.0341\\
NGC 5995     &&&&&5&0.0252&NGC 3347&x&&x&&3&0.0104\\
NGC 6552     &&&&&1&0.0265&NGC 1459&&&&&4&0.0139\\
NGC 7212     &M&&&&2&0.0266&NGC 2211&x&&&&-1&0.0067\\
NGC 7314     &&&&&4&0.0048&MRK 41&&x&&&1&0.0194\\
NGC 7674     &s&x&&&4&0.0289&NGC 6990&&&&&1&0.0320\\
NGC 7682     &&x&&x&2&0.0171&NGC 897&&&&&1&0.0159\\
\hline
P&0.5077&0.7723&0.2831&0.1978&&&&&&&&&\\
\hline
 \end{tabular} 
 \tablefoot{{\it(1)} Name as in Wu et al. (2011), {\it(2-5)} projected radial distance bin in $\kpc$, with x we mark all neighbors 
that are no more than two magnitudes fainter than the Seyfert or control sample galaxy, while with s the ones that are between 
two and three magnitudes fainter. We use M for mergers, * for multiple neighbors ($>3$), and 
$\star$ for objects with no directly comparable magnitude to the sample galaxy. 
{\it(6)} Hubble type T from -1 for S0 to 5 for Sc. 9 is for peculiar, {\it(7)} spectroscopic redshift.}
 \end{minipage}
 \end{table*}  

\newpage
%\begin{landscape}
\begin{table*}
\begin{minipage}{175mm}
\caption{Non-HBLR Sy2 and control sample}
\tabcolsep 5pt
\begin{tabular}{lccccrc|lccccrcc}
{\em Name} &{\em 50} & {\em 100}&{\em 150}&{\em 200}&  {\em T$\;\;\;$}&{\em z} &
{\em Name} &{\em 50} & {\em 100}&{\em 150}&{\em 200}&  {\em T$\;\;\;$}&{\em z} &\\
{\em (1)} &{\em (2)} & {\em (3)}&{\em (4)}&{\em (5)}&  {\em (6)$\;\;$}&  {\em (7)}&
{\em (1)} &{\em (2)} & {\em (3)}&{\em (4)}&{\em (5)}&  {\em (6)$\;\;$}&  {\em (7)}\\
\hline
&&&&&&&&&&&&\\
F01428--0404 &&&&&4--pec&0.0182&NGC 776&&x&s&&3&0.0164\\
ESO 428--G014&&xs&&&-1--pec&0.0057&NGC 1160&&&&&5&0.0084\\
NGC 1143     &Mx&x&&&pec&0.0282&NGC 1244&&&&&2--pec&0.0184\\ 
NGC 1144     &M&x&x&&pec&0.0282&NGC 1341&&&&&2&0.0063\\
NGC 1241     &x&&&&3&0.0135&NGC 1792&&&&x&4&0.0040\\
NGC 1320     &x&&&&1&0.0089&NGC 2785&&x&&&pec&0.0087\\
NGC 1358     &&&x&&0 &0.0134&NGC 3248&&&&&-1&0.0051\\
NGC 1386     &x&x&x&xx&-1&0.0029&NGC 3381&&&&&1--pec&0.0054\\
NGC 1667     &&&&x$\star$&5&0.0152&NGC 3462&&&&&-1&0.0215\\
NGC 1685     &&&x&xx&0&0.0152&NGC 3500&&&&&2&0.0116\\
NGC 3079     &&s&&&5 &0.0037&NGC 3760&&&&&0&0.0044\\
NGC 3281     &&&&x&2&0.0107&NGC 4082&s&&&&4 &0.0233\\
NGC 3362     &&&&&5&0.0277&NGC 4217&&&&&3&0.0034\\
NGC 3393     &&s&&&1&0.0125&NGC 4573&&&&&0&0.0099\\
NGC 3660     &&&&&4&0.0123&NGC 4601&&&x&x&0&0.0107\\
NGC 3982     &&x&xx&&3&0.0037&NGC 4665&&&&x&0 &0.0026\\
NGC 4117     &xx&&x&&-1&0.0031&NGC 4679&&&x&&5--pec&0.0155\\
NGC 4501     &&&&&3 &0.0076&NGC 4800&&&&&3  &0.0030\\
NGC 4941     &&&&x&2&0.0037&NGC 5134&&&&&3 &0.0059\\
NGC 5128     &&&&&-1--pec&0.0020&NGC 7773&&&&&4&0.0283\\
NGC 5135     &&&&&2&0.0137&NGC 5743&x&&&&3&0.0137\\
NGC 5194     &x&&&&4--pec&0.0015&NGC 5829&&&&&5&0.0188\\
NGC 5256     &M&&&&pec&0.0280&NGC 6030&s&&&&-1&0.0147\\
NGC 5283     &&&&&-1&0.0104&NGC 6403&s&&x&&-1&0.0163\\
NGC 5347     &&&&&2 &0.0080&NGC 7600&&&&& -1  &0.0116\\
NGC 5643     &&&&&5&0.0040&NGC 7683&&&&&-1&0.0124\\
NGC 5695     &&&&&3&0.0141&NGC 7779&x&&&x&0&0.0170\\
NGC 5728     &&&&&1&0.0094&NGC 429 &&x&x&&-1&0.0188\\
NGC 5929     &M&s&&&2--pec&0.0083&NGC 5383&s&&&&3--pec&0.0076\\
NGC 6300     &&&&&3&0.0037&NGC 1422&&xx&&x&2--pec&0.0055\\
NGC 6890     &&&&&3&0.0081&NGC 1463&&&&&1&0.0209\\
NGC 7130     &&&&&1--pec&0.0162&NGC 1511&&s&&&1--pec&0.0045\\
NGC 7172     &&xxxs&&&1--pec&0.0087&NGC 1964&&&&&3&0.0055\\
NGC 7496     &&&&&3&0.0055&NGC 3038&&&&x&3&0.0093\\
NGC 7582     &&xx&&x&2&0.0053&NGC 3188&x&&&&2&0.0260\\
NGC 7590     &x&x&&x&4 &0.0053&NGC 3600&&&&&1&0.0024\\
NGC 7672     &&x&&&3&0.0134&NGC 5233&&&&&2&0.0265\\
MRK 334      &M&&&&pec&0.0219&MRK 439 &&&&&1&0.0035\\
MRK 938      &M&&&&pec&0.0196&MRK 677 &&&&&3&0.0248\\
MRK 1066     &M&&&&-1&0.0120&MRK 1039&M&s&&&5 &0.0051\\
MRK 1361     &&&&&1&0.0226&MRK 1171&&&&x&5&0.0173\\
IC 5298      &&&&&1&0.0274&IC 5198 &&&&&1&0.0138\\
UGC 6100     &&&&&1&0.0295&UGC 6200&&&&&5&0.0129\\
\hline
$P_{null}$&0.0141&0.0168&0.0358&0.0640&&&&&&&&&\\
\hline
\end{tabular} 
\tablefoot{{\it(1)-(7)} as in Table 1.}
\end{minipage}
\end{table*}  
%\end{landscape} 

\end{document}